\documentclass[12pt]{article}

\input{epsf}
\usepackage[dvips]{graphicx}
\usepackage{cite}
\def\1ad{\mbox{\normalsize $^1$}}
\def\2ad{\mbox{\normalsize $^2$}}
\def\3ad{\mbox{\normalsize $^3$}}
\def\4ad{\mbox{\normalsize $^4$}}

\def\5ad{\mbox{\normalsize $^5$}}
\def\6ad{\mbox{\normalsize $^6$}}
\def\7ad{\mbox{\normalsize $^7$}}
\def\8ad{\mbox{\normalsize $^8$}}

\def\GE73{{}^{73}{\rm Ge}}
\def\gE76{{}^{76}{\rm Ge}}
\def\gEe74{{}^{74}{\rm Ge}}
\def\xe131{{}^{131}{\rm Xe}}
\def\i127{{}^{127}{\rm I}}

\def\dj{\hbox{d\kern-0.347em \vrule width 0.3em height 1.252ex depth
-1.21ex \kern 0.051em}}

\def\bun{{\overline u}_{\nu}}
\def\un{u_{\nu}}

\def\nuone{\nu^{(1)}}

\def\bone{B^{(1)}}

\newcommand{\be}{\begin{equation}}
\newcommand{\ee}{\end{equation}}
\newcommand{\ben}{\begin{equation*}}
\newcommand{\een}{\end{equation*}}
\newcommand{\ba}{\begin{eqnarray}}
\newcommand{\ea}{\end{eqnarray}}
\newcommand{\ban}{\begin{eqnarray*}}
\newcommand{\ean}{\end{eqnarray*}}
\newcommand{\brr}{\begin{array}}
\newcommand{\err}{\end{array}}
\newcommand{\bc}{\begin{center}}
\newcommand{\ec}{\end{center}}

\newcommand{\bea}{\begin{eqnarray}}
\newcommand{\eea}{\end{eqnarray}}
\newcommand{\bean}{\begin{eqnarray*}}
\newcommand{\eean}{\end{eqnarray*}}

\newcommand{\ie}{\mbox{\it i.e.~}}

\newcommand\lsim{\mathrel{\rlap{\lower4pt\hbox{\hskip1pt$\sim$}}
    \raise1pt\hbox{$<$}}}
\newcommand\gsim{\mathrel{\rlap{\lower4pt\hbox{\hskip1pt$\sim$}}
    \raise1pt\hbox{$>$}}}
\newcommand{\psla}{\not \! p}
\newcommand{\bfq}{{\bf q}}
\begin{document} 

\setcounter{page}{0}
\thispagestyle{empty}

\begin{flushright}
ANL-HEP-PR-02-054\\
EFI-02-94\\
FERMILAB-Pub-02/215-T\\
hep-ph/0209262
\end{flushright}

\vskip 8pt

\begin{center}
{\bf \Large {Elastic Scattering and Direct Detection \\[0.25cm] 
of Kaluza--Klein Dark Matter}}
\end{center}

\vskip 10pt

\begin{center}
{\large G\'eraldine Servant $^{a,b}$ and Tim M.P. Tait $^{a,c}$}
\end{center}

\vskip 20pt

\begin{center}
\centerline{$^{a}$ {\it High Energy Physics Division, Argonne National 
Laboratory, Argonne, IL 60439}}
\vskip 3pt
\centerline{$^{b}${\it Enrico Fermi Institute, University of Chicago, Chicago, 
IL 60637}}
\vskip 3pt
\centerline{$^{c}${\it Fermi National Accelerator Laboratory, P.O. Box 500, Batavia, IL 60510}}
\vskip .3cm
\centerline{\tt  servant@theory.uchicago.edu, tait@fnal.gov}
\end{center}

\vskip 13pt

\begin{abstract}
\vskip 3pt
\noindent

Recently a new dark matter candidate has been proposed as a consequence of 
universal compact extra dimensions. It was found that to 
account for cosmological observations, the masses of the first 
Kaluza-Klein modes (and thus the approximate size of the extra dimension) 
should be in the range 600-1200 GeV when the lightest Kaluza-Klein
particle (LKP) corresponds to the hypercharge boson and in the range 1 - 1.8 TeV
when it corresponds to a neutrino.  In this article, we compute the 
elastic scattering cross sections between Kaluza-Klein dark matter 
and nuclei both when the lightest Kaluza-Klein particle is
a KK mode of a weak gauge boson, and when it is a neutrino.  We include
nuclear form factor effects which 
are important to take into account due to the large LKP masses favored 
by estimates of the relic density. We present both differential and 
integrated rates for present and proposed Germanium, NaI and Xenon detectors.
Observable rates at current detectors are typically less than one event per
year, but the next generation of detectors can probe a significant fraction
of the relevant parameter space.

\end{abstract}

\vskip 13pt
\newpage
\section{Introduction}
 
Recently a new type of dark matter candidate has been proposed, 
consisting of stable Kaluza-Klein (KK) modes of ordinary standard 
model particles allowed to propagate in one or more compact extra dimensions
\cite{Servant:2002aq,Dienes:1998vg,Cheng:2002ab}.  
Under reasonably mild
assumptions about the nature of the UV completion, a five dimensional (5d)
theory with orbifold boundary conditions has a lightest Kaluza-Klein particle
(LKP) which is stable, and thus would be present as a cold relic in the
universe today.  If one further assumes that the LKP is a neutral, non-baryonic
particle, it has all of the properties required of a well-motivated
weakly interacting massive particle (WIMP), and what remains is to 
determine the relevant range of masses and other parameters
needed in order to correctly account 
for cosmological observations, and to determine the sensitivity of current 
and future dark matter searches to detect it either directly or indirectly.

Models with compact extra dimensions possess a very rich phenomenology and 
have attracted much attention lately. 
Fields which are permitted to propagate in the extra dimensions appear to a low
energy observer as a tower of increasingly massive particles, each with 
identical spin and charge.  The massive states are in fact modes of the fields
which carry (quantized) momentum in the extra dimensions, and thus the spacing
of the tower is roughly $1/R$, the inverse size of the extra dimensions.  
The ordinary 
particles of everyday experience are identified with the ``zero modes'', 
carrying zero momentum in the compact directions. Models with extra dimensions may be
classified according to which fields are allowed to propagate in extra dimensions. In brane
world models, only gravitational fields  can propagate in extra dimensions and therefore have
KK excitations. There are other interesting models where standard model gauge bosons live in
the bulk as well while standard model matter fermions are still confined on a 3-dimensional 
surface. And finally, there exist models --the so-called models with {\it Universal Extra
Dimensions} (UED) \cite{Appelquist:2000nn}-- in which all standard model fields have KK modes.
 These three classes of
models have very distinct phenomenology. UED models are the only models with extra dimensions
to have a stable KK particle because of a discrete symmetry (KK parity) which is remnant 
of the higher 
dimensional Poincar\'{e} invariance.  Under KK parity, all even mode-number particles 
(including the zero modes) are even, while the odd modes are odd.  This has
the desired consequence that the lightest odd mode must be stable.  If
this lightest KK Particle is neutral and weakly interacting, it
provides a cold dark matter candidate.

Having fermions in extra dimensions requires further assumptions about the
nature of the compact dimensions.  Generally, from the four dimensional 
point of view, a fermion in higher dimensions will not have chiral 
interactions, a necessary ingredient of any electroweak theory.  This may be
addressed in the UED context by imposing orbifold boundary conditions which 
project out the zero modes of the unwanted degrees of freedon responsible for vector-like
interactions.  For five dimensions, this can be accomplished by
identifying the coordinate of the extra dimension $y$ with $-y$, 
folding the circle $S^1$ onto the line $S^1 / Z_2$.
In six dimensions, starting from a 2-torus $T^2$, one can identify either
points related by a rotation of 180 degrees ($T^2 / Z_2$) or by 90 degrees
($T^2 / Z_4$).  One can similarly consider cases in even more dimensions,
but we will restrict our discussion, for simplicity, to the five or 
six dimensional cases.
Having imposed the orbifold in order to recover a chiral low energy theory,
it can be shown \cite{Georgi:2000ks} that the needed boundary conditions
imply that there are terms in the Lagrangian which live on the
fixed points of the orbifold transformation.  In five dimensions, these
are the points on the boundaries of the extra dimension.  These boundary
terms cannot be computed in terms of other parameters without knowing 
the UV completion of the theory. Consequently they must instead
be treated as parameters of the UED model. We expect that the masses
of the first level KK modes should be of order $1/R$, but will have corrections
from the boundary terms which will in general be different for different 
fields.

The most interesting cases for dark matter are when the boundary terms are
such that the LKP is a KK mode of either a neutrino, a neutral Higgs or a neutral 
weak gauge boson.  In this work we focus on the neutrino and gauge boson possibilities. 
Because of electroweak symmetry-breaking, the KK towers of the hypercharge
boson $B$ and the neutral $SU(2)$ boson $W_3$ mix.  The mass matrix for the
first level KK modes (in five dimensions) may be expressed in the
($\bone$, $W_3^{(1)}$) basis,
\bea
\left(
\begin{array}{cc}
\frac{1}{R^2} + \frac{1}{4}g_1^2 v^2 
+ \delta M_1^2 & \frac{1}{4}g_1 g_2 v^2 \\
\frac{1}{4}g_1 g_2 v^2 & \frac{1}{R^2} + 
\frac{1}{4}g_2^2 v^2 + \delta M_2^2
\end{array}
\right) ,
\eea
where $R$ is the size of the extra dimension, $v$ is the Higgs vacuum
expectation value, $g_1$ and $g_2$ are the gauge couplings, and
$\delta M_i^2$ are the boundary terms.  If the boundary terms are induced
radiatively, they should be proportional to the gauge couplings and $1/R^2$.
Thus, for $1/R \gg v$, the matrix is rather close to diagonal, and since
$g_1 < g_2$ we can expect that the lighter particle is well-approximated
as being entirely $\bone$.  Thus, in this case the LKP is a massive neutral
vector particle which couples to matter proportionally to 
$g_1$ times the hypercharge.

In Ref.~\cite{Servant:2002aq}, we determined the relic density for the LKP
when it is either a KK mode of a neutrino ($\nuone$) or of a neutral 
gauge boson ($\bone$).  As seen above, these represent
natural candidates when terms confined to the orbifold fixed points are
taken to arise radiatively\footnote{$\bone$ is indeed the LKP when assuming 
vanishing boundary terms at the cut-off scale. Note
 that this kind of prescription is similar to the choice of universal soft SUSY
 breaking masses at the GUT scale in susy models.} \cite{Cheng:2002iz}, as opposed to being present 
at tree-level \cite{Carena:2002me}. A variety of co-annihilation channels 
were included, with a range of mass splittings between the LKP and heavier
first tier KK modes, and the conclusion is that in order to correctly account
for the observed density of dark matter, the LKP masses should lie in the
ranges $600$ to $1200$ GeV for $\bone$ and $1000$ and $1800$ GeV for 
$\nuone$. We also noted that for a six dimensional (6D) 
orbifold $T^2/Z_2$, these mass ranges are lowered by 
approximately a factor of $\sqrt{2}$. Under our assumption of 
small boundary terms, the LKP mass 
corresponds to the inverse radius of the compact dimension and we expect all 
first level KK modes to have masses of this order.  The range relevant for
dark matter is particularly tantalizing because it lies just above the
current bounds from high energy colliders \cite{Appelquist:2000nn}\footnote{
The possibility of an extra dimension at a TeV was first examined 
in \cite{Antoniadis:1990ew}.}. Given 
the very large number of currently running or planned experiments devoted 
to both direct and indirect searches for WIMPs, the detectability of KK 
dark matter is an interesting question. Indirect detection issues have 
recently started to be investigated \cite{Cheng:2002ej,Hooper:2002gs,Bertone:2002ms}.

Direct detection of a weakly interacting massive particle typically involves
searching for the rare scattering of the WIMP with a nucleus in a detector.
As a result of the interaction, the nucleus recoils with some energy, which
can be read out as a signal \cite{Goodman:1984dc}.  The distribution of recoil 
energies is a function of the masses of the WIMP and the nucleus, and 
(because the scattering length for heavy WIMPs is typically of the same 
order as the size of the nucleus) the nuclear wave function.  The lightest 
supersymmetric particle (LSP) is a typical Majorana fermion WIMP with mass 
on the order of 100 GeV, and theoretical predictions for its interactions 
at modern dark matter detectors have reached a high level of sophistication.
In Ref.~\cite{Cheng:2002ej}, computations for the cross sections of $\bone$ 
scattering with nucleons were performed and prospects for direct detection
at present experiments were presented. In contrast with 
Supersymmetric WIMPs, predictions depend only on three 
parameters: The LKP mass, the
mass difference between the LKP and the KK quarks (assuming all flavors and
chiralities of first level KK quarks are degenerate in mass) 
and the ``zero mode'' Higgs mass. It is clear that present experiments can only probe
KK masses below 400 GeV as soon as the mass splitting between the LKP and KK quarks is
larger than five percents (as is found in \cite{Cheng:2002iz}). On the other hand, masses
below 300 GeV are already excluded by collider constraints\cite{Appelquist:2000nn}. And 
in any case, masses
below 400 GeV are in conflict with the mass range predicted from our relic density
calculation \cite{Servant:2002aq}. Therefore, we wish to investigate detection 
prospects for masses above
400 GeV and ask: Which planned experiment will be able to probe the relevant
parameter space of the LKP? To answer this question, we need to go
beyond the wimp-nucleon cross section calculation and compute
the event rate for a given detector. This requires to include the nuclear
form factor. In the current article we expand
upon the results of Ref.~\cite{Cheng:2002ej}, deriving realistic estimates for 
event rates at modern dark matter detectors, including nuclear wave function 
effects and examining differential rates in the nuclear recoil energy as well as 
integrated ones. We find that the event 
rate is somewhat smaller than for the usual LSP neutralino WIMP with
mass around 100 GeV and that in order to see at least several events 
per year, heavy ($>$ 100 kg) detectors are needed. 

This article is organized as follows.
In section II we review the kinematics of direct detection of WIMPs. 
In section III, we present the analysis in the case where the LKP is the 
first KK state of the neutrino, finding that it should most likely have
already been observed by CDMS or EDELWEISS, and thus is excluded.  
Section IV is devoted 
to the more interesting case of $\bone$. Our predictions for the differential
and integrated event rates expected in Germanium, Sodium--Iodide and 
Xenon detectors are presented in section V.  We reserve section VI for our
conclusions and outlook.

\section{Kinematics of WIMP Detection}
\label{sec:reminder}

In this section we briefly review the general kinematics of WIMP-nucleus
scattering.  The number of events per unit time and per unit detector mass is,
\be
\label{rate}
dR=\frac{\rho}{m M}\frac{d\sigma}{d|\bfq|^2}d|\bfq|^2 v f(v) dv ,
\ee
where $m$ is the WIMP mass and $\rho$ its mass density in our solar 
system\footnote{In our numerical results, we identify $\rho$ with the local 
(in our galaxy) dark matter density of canonical value $\rho \sim 0.3 \ 
 {\rm  GeV}/{\rm cm}^3$ \cite{Gates:1995dw}
\ie we assume a homogeneous halo of our galaxy. This assumption is not obvious. 
In addition, estimates of the local density of dark matter in our galaxy is subject to
considerable uncertainty and model-dependance. We should then keep in mind that local
over- or underdensities could easily change the expected detection rate by a significant
amount.},
and $M$ is the mass of the target nucleus.
$f(v)$ is the distribution of WIMP velocities relative to the detector, 
$\mu \equiv m M/(m+M) $ is the reduced mass, $q^\mu$ is the momentum 
transfer four-vector whose magnitude is $|\bfq|^2=2 \mu^2 v^2(1-\cos \theta)$ 
in terms of  $\theta$, the scattering angle in the center of momentum frame. 
$|\bfq|^2$ is related to the recoil kinetic energy $E_r$ deposited in the 
detector (in the lab frame) by $E_r=|\bfq|^2/2M$. For $m \gg M$ as is the
case for LKP WIMPs with masses of order 1 TeV, $E_r$ is typically 
30--50 keV depending on the nucleus target (but it can be much larger for 
WIMPs with velocities close to the galactic escape velocity).  
Eq.~(\ref{rate}) may be thus rewritten
\be
\frac{dR}{dE_r}=\frac{2\rho}{m }\frac{d\sigma}{d|\bfq|^2} v f(v) dv ,
\ee
in which $|\bfq|^2$ should be regarded as a function of $E_r$ as 
indicated above.
The differential cross section can be expressed in terms of the cross 
section at zero momentum transfer $\sigma_0$ times a nuclear form 
factor \cite{Jungman:1995df},
\be
\frac{d\sigma}{d|\bfq|^2}=\frac{\sigma_0}{4 \mu^2 v^2}F^2(|\bfq|) ,
\ee
where $F^2(|\bfq|)$ is a function normalized to one at $|\bfq|^2=0$
which includes all relevant nuclear effects and must be determined either
directly from measurements of nuclear properties or estimated from a 
nuclear model, and $\sigma_0$ contains
the model-dependent factors for a specific WIMP.
The rate is obtained by integrating over all possible incoming 
velocities of the WIMP:
\be
\frac{dR}{dE_r}=\frac{\sigma_0\rho}{2m \mu^2}F^2(|\bfq|) 
\int_{v_{min}}^{v_{max}}\frac{f(v)}{v} dv ,
\ee
where $v_{max} \simeq 650$ km/s, the galactic escape velocity. 
To determine $v_{min}$ we use the relation between the WIMP energy $E$ 
and the recoil energy $E_r$
\bea
E&=&  \frac{2E_r}{1-\cos \theta}\frac{(m+M)^2}{4m M}\rightarrow
E_{min}=E_r\frac{(m+M)^2}{4m M}  ,\\
v_{min}&=&(2 E_{min}/m)^{1/2}=\sqrt{\frac{E_r M}{2 \mu^2}} .
\eea
Assuming a Maxwellian velocity distribution for the WIMPs and including the 
motion of the Sun and the Earth one obtains \cite{Jungman:1995df},
\be
\int_{v_{min}}^{\infty} \frac{f(v)}{v} dv=
\frac{1}{2v_E}\left[\mbox{erf}\left(\frac{v_{min}+v_E}{v_0}\right)
-\mbox{erf}\left(\frac{v_{min}-v_E}{v_0}\right)\right] ,
\ee
where $v_E$ is the relative motion of the observer on the Earth to the sun
(and thus shows an annual modulation),
and $v_0$ is the mean relative velocity of the sun relative to the 
galactic center\footnote{Together with $\rho$, note that $v_0$ is another crucial quantity in both direct
and indirect methods of dark matter detection which is subject to significant uncertainties.}.
Thus, the final formula for the measured differential event rate is,
\be
\label{eq:dRdE}
\frac{dR}{dE_r}=\frac{\sigma_0\rho}{4v_Em \mu^2}F^2(|\bfq|)
\left[\mbox{erf}\left(\frac{v_{min}+v_E}{v_0}\right)
-\mbox{erf}\left(\frac{v_{min}-v_E}{v_0}\right)\right] .
\ee
The total event rates per unit detector mass and per unit time will depend
on the range of energies to which the detector is sensitive.  Thus, the
actual observed rate, modulo experimental efficiencies, will be 
given by $dR / dE_r$ integrated over the appropriate range of energy 
for a given experiment.

Our task will now be to compute $\sigma_0$ and to combine it with the correct
form factor $F^2(|\bfq|)$ in cases where the WIMP is $\nuone$ or 
$\bone$. To compute $\sigma_0$, we must evaluate the effective WIMP
interaction with nuclei by evaluating the matrix elements of the nucleon 
operators in a nuclear state.  This in turn is determined from WIMP 
interactions with quarks and gluons evaluated in nucleon states.
Traditionally, one differentiates between two very different types of
WIMP-nucleon interactions- {\it spin-dependent } interactions and 
{\it scalar} interactions. 

Scalar interactions are coherent between nucleons in the nucleus, 
and the form factor is thus the Fourier transform of the nucleon density. 
The commonly used form (identical, in the limit of low momentum transfer, to 
the one derived from a {\it Woods--Saxon} parametrization of the nuclear 
density \cite{Jungman:1995df,Engel:bf}) is :
\be
\label{eq:SFF}
F^2(|\bfq|)=\left(\frac{3j_1(qR_1))}{qR_1}\right)^2e^{-(qs)^2} ,
\ee
where $R_1=\sqrt{R^2-5s^2}$ and $R\sim 1.2$ fm A$^{1/3}$ with $A$ the nuclear mass number, 
$s\sim 1$ fm and $j_1$ is a spherical Bessel function,
\bea
j_1 (q r_n) & = & \frac{ \sin [ q r_n ] - q r_n \cos [ q r_n ]}{(q r_n)^2} .
\eea

An axial-vector interaction leads to interactions between the WIMP spin and
nucleon spin.  In this case one must evaluate the matrix elements of 
nucleon spin operators in the nuclear state. The form factor is 
typically written as \cite{Jungman:1995df,Engel:bf} ,
\be
\label{eq:spindepFF}
F^2(|\bfq|)=\frac{S(|\bfq|)}{S(0)} ,
\ee
where, 
\bea
S(|\bfq|)&=&a_0^2 S_{00}(|\bfq|)+a_1^2 
S_{11}(|\bfq|)+a_0a_1S_{01}(|\bfq|) ,\\
a_0&=&a_p+a_n, \ \ a_1=a_p-a_n, 
\eea
where the first term is the iso-scalar contribution, the second one is the 
iso-vector contribution and the last one is the interference term between 
the two.
The $S_{ij}$ are obtained from nuclear calculations. $a_p$ and $a_n$ reflect 
the spin-dependent WIMP interactions and 
average spins for neutrons and protons in the nucleus and will be
defined below.

\section{Direct Detection of $\nuone$}
\label{sec:neutrino}
\begin{figure}[h]
\begin{center}
\includegraphics[height=3cm]{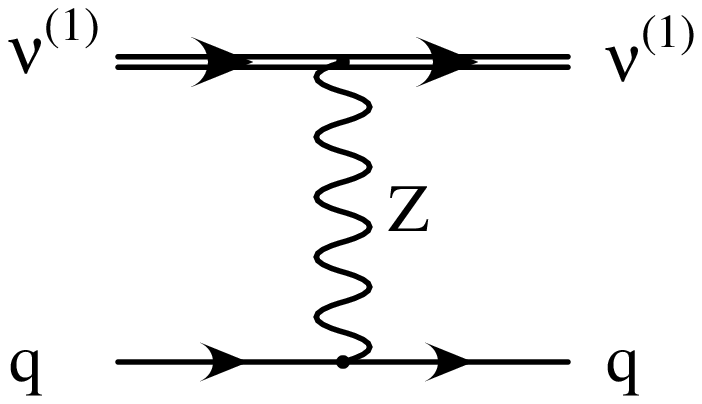}
\end{center}
\caption[]{Leading Feynman graph for effective $\nuone$-quark 
scattering through the exchange of a zero-mode Z gauge boson.}
\label{fig:neutrino}
\end{figure}

For our first example, we consider the KK neutrino, $\nuone$.  This is almost
a case which has been considered previously \cite{Goodman:1984dc}, the only
difference being that the KK neutrino has vector-like weak interactions.
In the non relativistic limit where $q^2 \ll m_Z^2$, 
we have an effective four-fermion contact interaction (Fig.~\ref{fig:neutrino}),
\be
\frac{-i e^2}{4 \sin^2 \theta_W m^2_W} \left[ \bun\gamma^{\mu}\un \right] 
\left( (g^q_R+g^q_L)
\left[ \overline{q}(x) \gamma_{\mu} {q}(x) \right]
+ (g^q_R-g^q_L) 
\left[ \overline{q}(x) \gamma_{\mu} \gamma_5 {q}(x) \right] 
\right),
\ee
where we have explicitly included the $Z$ couplings to $\nuone$,
\be
g^{\nuone}_R=g^{\nuone}_L=\frac{e}{2 \sin \theta_W \cos \theta_W},
\ee
and the $g^q_L$ and $g^q_R$ are the left- and right-handed quark interactions
with the $Z$ boson,
\bea
g^q & = & T^q_3 - Q_q \sin^2 \theta_W .
\eea
Thus we see that the effective interaction includes both a coupling to
the vector and the axial vector quark currents.  When evaluating the
WIMP-nucleon cross section, this will be summed over all flavors of quarks
and will involve matrix elements $\langle \overline{q} \gamma_{\mu} q \rangle$
and $\langle \overline{q} \gamma_{\mu}\gamma_5 q \rangle$, where the
expectation values are to be understood as refering to nucleon states.


The WIMPs are highly non-relativistic, and thus only the time-component of the
vector $\bun \gamma^{\mu} \un$ is appreciable.  However, the expectation value
$\langle \overline{q} \gamma_0 \gamma_5 q \rangle \simeq 0$
\cite{Engel:bf}, and we are left with only the time-component of the
vector interaction.  This illustrates the predominant difference between
$\nuone$ and a typical massive Dirac neutrino WIMP - the absence of 
spin-dependent interactions.  However, since the spin-dependent contribution
is usually sub-dominant to the scalar interaction, the resulting cross
sections remain comparable.

At the quark level, the effective interaction has the form,
\bea
b_q  \left[ \bun\gamma^{\mu}\un \right] 
\left[ \overline{q}(x) \gamma_{\mu} {q}(x) \right],
\eea
where,
\bea
b_q & = & \frac{e^2}{4 \sin^2 \theta_W m^2_W}
\left[ T^q_3 - 2 Q_q \sin^2 \theta_W \right] .
\eea
The matrix element 
$\langle \overline{q} \gamma^0 q \rangle = \langle q^\dagger q \rangle$
simply counts valence quarks in the nucleon, and so the nucleon WIMP
couplings are,
\bea
b_p &=& 2 b_u + b_d = \frac{G_F}{\sqrt{2}} 
\left(1 - 4 \sin^2 \theta_W \right) , \\
b_n &=& 2 b_d + b_u = -\frac{G_F}{\sqrt{2}} ,
\eea
for the proton and neutron, respectively.  The numerical accident that
$\sin^2 \theta_W \simeq 1/4$ renders the coupling to protons very small.
The vector interactions are coherent, and thus we have for the WIMP-nucleus
coupling, $b_N= Z b_p + (A-Z)b_n$.  Thus,
\bea
\sigma_0 &=&
\frac{\mu^2 G_F^2}{2 \pi}
\left[ (1-4\sin^2\theta_W)Z - (A-Z)\right]^2
\eea
and the form factor entering in the differential cross section 
$d\sigma/d\bfq^2$ is given by (\ref{eq:SFF}).

It is well-known that the mass of Dirac neutrinos is strongly constrained 
by elastic scattering experiments such as CDMS \cite{Abrams:2002nb}
and EDELWEISS \cite{Benoit:2001zu}.
The exclusion plots are presented in the $m$-$\sigma_n$ plane where 
$m$ is the mass of the dark matter candidate and $\sigma_n$ is the 
scattering cross section per nucleon. It is related to $\sigma_0 $ by,
\bea
\sigma_n &=& \sigma_0 \frac{m_n^2}{\mu^2 A^2}
\eea
$m_n$ being the mass of the nucleon.
For $\GE73$, and $m \gg M$, we find 
$\sigma_n\sim 2\times 10^{-39}$ cm$^2 \sim 2\times 10^{-3}$ pb.  Given that
CDMS and EDELWEISS did not see any events, a WIMP with this cross section
must have a mass $\gsim 50$ TeV.  This means that in order to have escaped
detection, $\nuone$ would have to have masses more than ten times larger than
the range of masses for which result in the correct dark matter relic density.
While one might imagine that coannihilation in various channels could 
push up the favored $\nuone$ masses by a few TeV , it seems unlikely that
the relic density calculation could favor masses above 10 TeV. 

To conclude this section, the KK neutrino seems to be ruled out as a 
dark matter candidate at least in the minimal UED model where the mass 
window prediction from the relic density calculation is in conflict with 
direct detection experiments. Let us therefore now concentrate on the 
$\bone$ LKP candidate.

\section{Direct Detection of $\bone$}
\label{sec:photon}

\begin{figure}[th]
\begin{center}
\includegraphics[height=3cm]{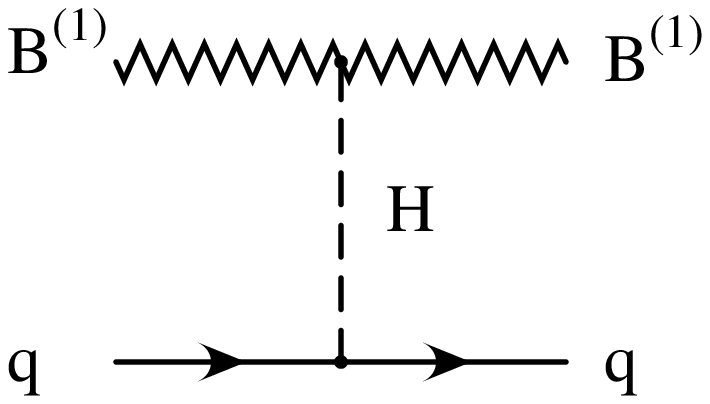}
\end{center}
\caption[]{Leading Feynman graph for effective $\bone$-quark scattering through
the exchange of a zero-mode Higgs boson.}
\label{fig:higgs}
\end{figure}

$\bone$ can interact elastically with a quark by exchanging a KK quark 
in the $s$- and $t$-channel or by $t$-channel Higgs exchange. 
The amplitude for scattering between quarks and $\bone$ mediated by 
Higgs exchange (Figure~\ref{fig:higgs}) is,
\be
{\cal M}_h= -i \gamma_q  \: 
\epsilon^*_\mu (p^\prime_B) \, \epsilon^\mu (p_B) \: \:
\left[ \bar{q}(x) \, q(x) \right] \ , \ \ \ \ 
\gamma_q=\frac{g_1^2}{2}\frac{m_q}{m_h^2} 
\ee 
where $\epsilon^\mu$ are the $\bone$ polarization vectors,
$q(x)$ is a quark field and there are separate couplings $\gamma_q$ for each
flavor of quark.  $g_1$ is the hypercharge coupling, and $Y_h= 1/2$ has been
explicitly included in the result.
We have taken the non-relativistic (NR) limit for the 
WIMPs in which we are justified in dropping tiny terms of order 
$(p_B - p_B^\prime)^2 / m^2_h$.  The factor of $m_q$ in $\gamma_q$ is a
direct consequence of the fact that zero mode quark masses result from the 
quark couplings to Higgs, after electroweak symmetry breaking.

\begin{figure}[th]
\begin{center}
\includegraphics[height=3cm]{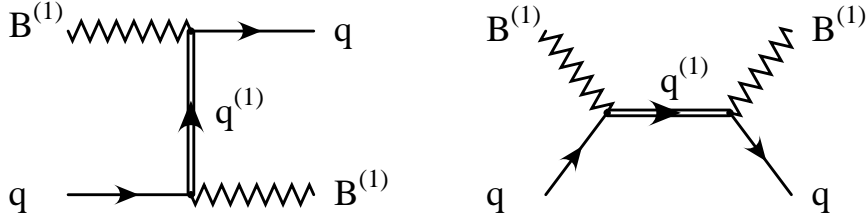}
\end{center}
\caption[]{Leading Feynman graphs for effective $\bone$-quark scattering 
through the exchange of a KK quark.  Both $q^1_L$ and $q^1_R$ should be
understood in each graph.}
\label{fig:quark}
\end{figure}

We now consider the KK quark exchange, with Feynman diagrams shown in
Figure~\ref{fig:quark}. Recall the coupling $\bone$-$q^{(1)}_{R(L)}$-$q$ 
involves a right (left)-handed projector and both $q^{(1)}_R$ and $q^{(1)}_L$ 
can be exchanged and will typically have somewhat different masses.  
Thus each Feynman graph of Figure~\ref{fig:quark} is actually two 
separate graphs with $q^{(1)}_L$ and $q^{(1)}_R$ exchanged.
The amplitudes corresponding to the two diagrams of Figure~\ref{fig:quark} 
are:
\bea
{\cal M}_1^{R/L}&=& -i(g_1 Y_{R/L})^2 \left[
\overline{q}(x) \gamma^{\nu}P_{R/L}
\frac{(\psla_q-\psla_{B^{\prime}}+m_{q^{(1)}})}
{(p_q-p_{B^{\prime}})^2-m^2_{q^{(1)}_{R/L}}}
\gamma^{\mu} P_{R/L} q(x) \right]
\epsilon^*_{\mu}(p^\prime_B) \epsilon_{\nu}(p_{B}), \hspace*{0.5cm} \\
{\cal M}_2^{R/L}&=& -i(g_1 Y_{R/L})^2 \left[
\overline{q}(x) \gamma^{\mu}P_{R/L}
\frac{(\psla_q+\psla_{B}+m_{q^{(1)}})}
{(p_q+p_B)^2-m^2_{q^{(1)}_{R/L}}}
\gamma^{\nu} P_{R/L} q(x) \right]
\epsilon^*_{\mu}(p_{B^{\prime}}) \epsilon_{\nu}(p_B) ,\hspace*{0.5cm}
\eea
where $Y_{R/L}$ are the hypercharges for the right- and left-chiral
quark $q$.
In the nonrelativistic limit
$p_B\approx p_{B^{\prime}}\approx(m_{\bone},0,0,0)$
so that ${\cal M}^{R/L}_q={\cal M}_1^{R/L}+{\cal M}_2^{R/L}$ can be rewritten:
\bea
{\cal M}_q^{R/L}&=& -ig_1^2 {Y^2_{q_{R/L}}} 
\epsilon^*_{\mu}(p_{B^{\prime}}) \ \epsilon_{\nu} (p_B)\times \\
&& \overline{q}(x) 
\left\{  \frac{(E_q+m_{\bone})\gamma^{\mu}\gamma^0\gamma^{\nu} }
{(m_{\bone}+E_q)^2-m^2_{q^{(1)}}}          
+\frac{(E_q-m_{\bone})\gamma^{\nu}\gamma^0\gamma^{\mu}}
{(m_{\bone}-E_q)^2-m^2_{q^{(1)}}} \right\}P_{R/L} 
\ q(x) \ ,
\nonumber
\eea
where we have neglected the 3-momentum of the quarks in the nucleon and 
written $p_q\approx
(E_q,0,0,0)$. $E_q<1 \ {\rm GeV}\ll M_{\bone}$ is the energy of a bound quark in the nucleon.
We now expand this expression up to linear order in 
$x=E_q m_{\bone}/(m^2_{\bone}-m^2_{{q^{(1)}}})$ to obtain
\be
{\cal M}_q^{R/L}= -i \epsilon^*_{\mu}(p_{B^{\prime}}) \ \epsilon_{\nu}(p_B) 
 \overline{q}(x) 
\left[S_q E^{\mu \nu} +A_q {\tilde{E}}^{\mu \nu}\right] P_{R/L} \ q(x) \ ,
\ee
where the coefficients $S_q $ (scalar contribution) and $A_q$ 
(spin-dependent contribution) are defined in equations (\ref{coeff}) and
\be
E^{\mu \nu} \equiv \gamma^{\mu}\gamma^0\gamma^{\nu}+
\gamma^{\nu}\gamma^0\gamma^{\mu} \ , \ \ 
{\tilde{E}}^{\mu \nu} \equiv \gamma^{\mu}\gamma^0\gamma^{\nu}-
\gamma^{\nu}\gamma^0\gamma^{\mu}=
2i\epsilon^{0\mu\nu\rho} \gamma_{\rho}\gamma_5 .
\ee
In the non relativistic limit $E^{\mu \nu}$ leads to scalar interactions
whereas ${\tilde{E}}^{\mu \nu}$ leads to spin-dependent interactions.

We will assume that all flavors and chiralities of 
first level KK quarks are equal and parameterize their masses by 
$\Delta = (m_{q^{(1)}} - m_{\bone}) / m_{\bone}$.
Summing  ${\cal M}_h$ and ${\cal M}^R_q + {\cal M}^L_q$ we obtain:
\be
\langle {\cal M}\rangle=-i \epsilon^*_{\mu}(p_B) \ 
\epsilon_{\nu}(p_{B^{\prime}})
\left[  (\gamma_q+S_q)g^{\mu\nu} \langle \overline{q} q \rangle 
+ A_q \langle \overline{q} {\tilde{E}}^{\mu \nu} q \rangle \right] ,
\ee
\be
\left|\langle {\cal M}_q\rangle\right|^2=
4(\gamma_q+S_q)^2 |\langle\overline{q} q \rangle|^2 + 
2 A_q^2|\langle\overline{q} \gamma^k\gamma_5 q\rangle|^2 ,
\ee
with,
\be
\langle\overline{q} q\rangle=\frac{m_p}{m_q}f^p_{T_q}.
\ee
We sum over the different quark contributions to obtain the matrix element 
in a nucleon state. At this stage, as a first-order evaluation, we will make the 
assumption $E_q\approx m_q$. We recognize
that the assumption that the light quarks in the nucleon are on-shell is questionable and that
a more accurate treatment would be desirable\footnote{
Note that this problem did not arise in the elastic scattering of a neutralino because 
neutralinos can only exchange bosons when interacting with spin-1/2 quarks. On the other hand,
 our bosonic dark matter candidate can exchange fermions with the quarks. This interaction
 involves fermion propagator and therefore involves the energy, and not only the mass, of the
 quark. Possibly the use of a constituent mass might be more appropriate in this case.}.

For the spin matrix element only the light quarks
$u,d,s$ contribute while for the scalar matrix elements there are also
contributions from heavy quarks $c,b,t$ \cite{Shifman:zn}:
\be
f_{p,n}^{\bone}=m_{p,n}\sum_q\frac{\gamma_q+S_q}{m_q}f^{p,n}_{T_q} .
\ee
Because of this distinction, we drop terms in $A_q$ which are 
proportional to any power of the zero mode quark 
mass since these are negligible. 
Note that under our assumption both $\gamma_q$ and $S_q$ are proportional to
 the quark mass $m_q$. 
Thus, given the normalization of the matrix elements $f_{T_q}$, each flavor
contributes to the scalar interaction proportionally to its contribution to 
the nucleon mass.  For heavy quarks $q=c,b,t$, the contribution should in fact 
be considered to be induced by the {\em gluon} content of the nucleon, with
the heavy quark legs closed to form a loop.  In the Higgs exchange case, the 
mapping from the tree graph with heavy quark external legs to the loop graph
with external gluons is straight-forwardly handled by the formalism of
\cite{Shifman:zn}.  For the KK quark graph, as emphasized in
\cite{Drees:1993bu}, this mapping is generally unreliable because of the 
presence of the heavy KK quark in the loop with mass $\sim m_{\bone}$.  Thus,
we include $q=c,b,t$ in $\gamma_q$, but to be conservative not in $S_q$.
From a practical point of view, the loop-suppression renders the contribution
from the heavy quarks irrelevant compared to the strange quark contribution,
so the final results are insensitive to this choice of procedure.

The coefficients $A_q$ and $S_q$ may be extracted from the matrix elements,
\bea
\label{coeff}
A_q&=&\frac{g_1^2(Y_{q_L}^2+Y^2_{q_R})m_{\bone}}
{(m^2_{\bone}-m^2_{q^{(1)}})} , \\
S_q&=&-E_q\frac{g_1^2(Y_{q_L}^2+Y^2_{q_R})}{(m^2_{\bone}-m^2_{q^{(1)}})^2}
(m^2_{\bone}+m^2_{q^{(1)}}) .
\nonumber
\eea
The total amplitude squared in a nucleus state reads:
\be
\left|\langle {\cal M}\rangle\right|^2=\left|\langle {\cal M}\rangle\right|^2_{scalar}+
\left|\langle {\cal M}\rangle\right|^2_{spin} ,
\ee
where,
\bea
\left|\langle {\cal M}\rangle\right|^2_{scalar}&=&4 m_N^2(Zf_p^{\bone}+(A-Z)f_n^{\bone})^2
\times F^2_{sc}(|\bfq|) , \\
\left|\langle {\cal M}\rangle\right|^2_{spin}&=&\frac{32}{3}g_1^4
\frac{m^2_{\bone}m_N^2}{(m^2_{\bone}-m^2_{q^{(1)}})^2}\Lambda^2J(J+1)
\times F^2_{sp}(|\bfq|) ,
\eea
so that the corresponding cross sections at zero momentum transfer, 
$\sigma_0$ are,
\bea
\sigma_0^{scalar}&=&\frac{m_N^2}{4\pi(m_{\bone}+m_N)^2}(Zf_p^{\bone}+(A-Z)f_n^{\bone})^2, \\
\sigma_0^{spin}&=&\frac{2}{3 \pi}\mu^2 g_1^4\frac{\Lambda^2 J(J+1)}{(m_{\bone}^2-m^2_{q^{(1)}})^2}.
\eea
where,
\be
\Lambda=\frac{a_p\langle S_p\rangle+a_n\langle S_n\rangle }{J},
\ee
and,
\be
a_{p(n)}=\sum_{u,d,s}(Y_{q_L}^2+Y_{q_R}^2) \Delta^{p(n)} q .
\ee{
Using $\Delta^{(n)} u=\Delta^{(p)} d \equiv \Delta d$, $\Delta^{(n)} d=\Delta^{(p)} u\equiv\Delta u$ 
and $\Delta^{(n)} s=\Delta^{(p)}s\equiv \Delta s$, we have,
\bea
a_p&=&\frac{17}{36}\Delta u + \frac{5}{36}(\Delta d+ \Delta  s), \\
a_n&=&\frac{17}{36}\Delta d + \frac{5}{36}(\Delta u+ \Delta  s),
\eea
where
$\Delta u =0.78 \pm 0.02$, \  $\Delta d= -0.48\pm 0.02$, \  
and $\Delta s=  -0.15\pm 0.02$ \cite{Ellis:2000ds}.
For the spin form factors we will also need the iso-singlet and iso-vector
combinations,
\bea
a_0&=&\frac{11}{18}(\Delta u+\Delta d)+\frac{5}{18}\Delta s, \\
a_1&=&\frac{1}{3}(\Delta u-\Delta d).
\eea
For Germanium and Iodine nuclei we have, assuming the WIMP mass is 1 TeV, 
$\Delta=(m_{q^{(1)}}-m_{\bone})/ m_{\bone}=15 \%$, and
the Higgs mass is $m_h=120$ GeV,
\bea
\mbox{Germanium}: \sigma_0^{scalar}=1.6 \times 10^{-3} \ {\rm pb}, 
\ \ \sigma_0^{spin}=1.4 \times 10^{-4} \ {\rm pb} , \\
\mbox{Iodine} \ \ \ \ \ \  : \sigma_0^{scalar}=1.34 \times 10^{-2} \ {\rm pb}, 
\ \ \sigma_0^{spin}=1.8 \times 10^{-3} {\rm pb} .
\eea
These values are somewhat smaller than what one would typically expect for 
neutralino-nucleus elastic scattering. In that case, 
one finds \cite{Jungman:1995df},
\bea
\sigma_{0,\chi}^{scalar}& \simeq &
\frac{4 \mu^2}{\pi}(Zf_p^{\chi}+(A-Z)f_n^{\chi})^2 \ \ \ , \ \ \ \\ 
\sigma_{0,\chi}^{spin}& \simeq &
\frac{32}{\pi}G_F^2\mu^2 \Lambda^2J(J+1) ,
\eea
so that,
\be
\frac{\sigma_{0,\bone}^{scalar}}{\sigma_{0,\chi}^{scalar}}
\sim\frac{1}{16m_{\bone}^2}\left(\frac{f_p^{\bone}}{f_p^{\chi}}\right)^2 .
\ee
Using $f_p=m_p\sum f_{T_q} f_q /m_q$ where $f_q^{\bone}\sim \gamma_q$ 
and $f_q^{\chi}\sim g_2 T_{h00} \ h_{hqq}/2m_h^2$ (note that  $f_q^{\bone}$ and 
$f_q^{\chi}$ have different dimensions) where 
$T_{h00}$ and $h_{hqq}$ are Higgs-neutralino-neutralino and Higgs-quark-quark
Yukawa couplings (which can be found, for instance, in 
Ref.~\cite{Jungman:1995df}), we have,
\be
\frac{\sigma_{0,\bone}^{scalar}}{\sigma_{0,\chi}^{scalar}} \sim
\left(\frac{g_1}{g_2}\right)^4\left(\frac{m_W}{m_{\bone}}\right)^2
\sim 10^{-3},
\ee
Therefore we expect $\sigma_{0,\chi}^{scalar}$ to be smaller than 
$\sigma_{0,\bone}^{scalar}$, however the ratio generally depends on 
the precise neutralino couplings, which are complicated functions of SUSY
parameter space.  We now compare spin-dependent cross sections:
\be
\frac{\sigma_{0,\bone}^{spin}}{\sigma_{0,\chi}^{spin}}\propto \frac{g_1^4}{48}
\left(\frac{a_p^{\bone}}{a_p^{\chi}}\right)^2
\frac{1}{G_F^2(m_{\bone}-m_{q^{(1)}})^2} \sim
\left(\frac{g_1}{g_2}\right)^4\frac{m_W^4}{(m_{\bone}^2-m^2_{q^{(1)}})^2}.
\ee
We again have a large suppression factor due to the large WIMP mass unless $ m_{q^{(1)}}$ is 
nearly degenerate with $m_{\bone}$.

\begin{figure}[th]
\label{fig:wimp-nucleon}
\begin{center}
\includegraphics[height=8cm]{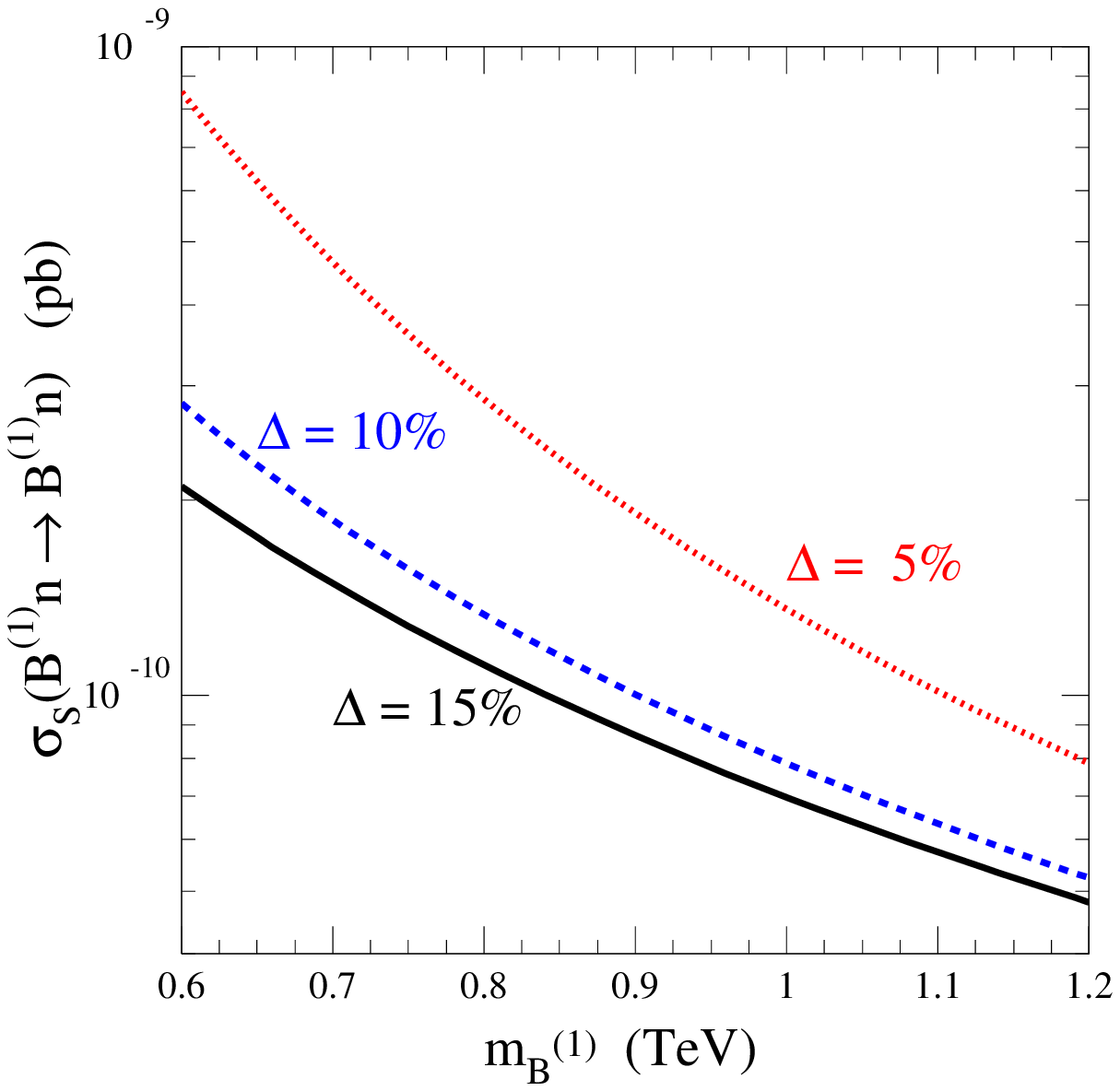} \hspace*{-1cm}
\includegraphics[height=8cm]{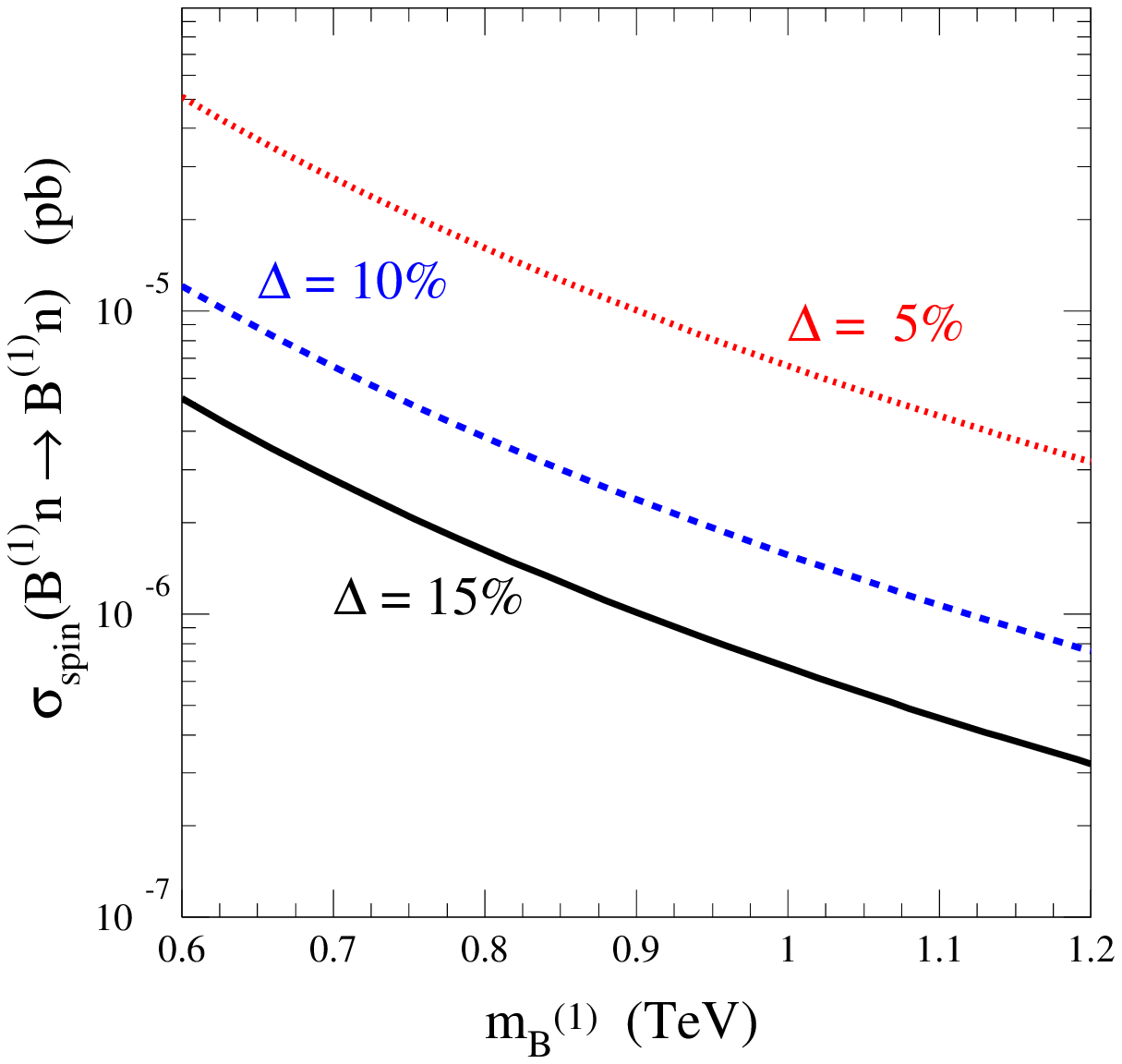}
\end{center}
\caption[]{Spin-dependent and spin-independent WIMP-nucleon cross sections 
as a function of the WIMP mass for (top to bottom)
$\Delta=(m_{q^{(1)}}-m_{\bone})/m_{\bone}=5,10,15 \%$ and $m_h=120$ GeV.}
\end{figure}

It is common for dark matter search experiments to express their constraints
in terms of effective WIMP-nucleon cross sections,
\bea
\sigma^{spin}_{p,n} &=& \frac{g_1^4}{2\pi}\frac{\mu_{p,n}^2a_{p,n}^2}
{(m_{\bone}^2-m^2_{q^{(1)}})^2}, \\
\sigma^{scalar}_{p,n} &=& \sigma_0\frac{m^2_{p,n}}{\mu^2}\frac{1}{A^2} .
\eea
For 1 TeV WIMP mass, typical values are 
$\sigma^{scalar}_{p,n}\sim 10^{-10}$ pb and 
$\sigma^{spin}_{p,n} \sim 10^{-6}$ pb.  
(For comparison, nucleon-neutralino cross sections are in the range 
$10^{-12}-10^{-6}$ pb for scalar interactions and $10^{-9}-10^{-4}$ pb 
for spin-dependent interactions).  
>From Fig.~\ref{fig:wimp-nucleon} we see that the cross sections may
vary upward by about one order of magnitude 
if $m_{\bone}$ is at the lower end of its favored range, 600 GeV, 
and by two orders of magnitude if in addition $\bone$ and ${q^{(1)}}$ are
more degenerate, $\Delta\sim 5 \%$.  The dependence on the 
zero-mode Higgs mass is presented in 
Fig.~\ref{fig:nucleonmh}.  Note that theories
in which the top and/or bottom quarks propagate in extra dimensions
\cite{Arkani-Hamed:2000hv} generically have additional contributions to 
electroweak observables through the oblique parameters $S$ and $T$
\cite{Appelquist:2000nn}, and thus the preference in the precision 
electroweak data for a light SM-like Higgs may be misleading in theories 
with universal extra dimensions.  Thus, we consider a wider range of
Higgs masses than one would naively expect from the electroweak fits.
Finally, in Fig.~\ref{fig:SIversusSD}, we show a scatter plot of
spin-dependent and spin-independent cross sections, varying
600 GeV $\leq m_{\bone} \leq$ 1200 GeV, $5\% \leq \Delta \leq 15\%$,
and 100 GeV $\leq m_h \leq 200$ GeV.

In any case, these cross sections are below the reach of any currently 
running experiment. However, larger mass detectors composed of heavier
nuclei and improved efficiencies will most likely change this situation
in the foreseeable future.
Since precise event rates will depend on experimental issues such as 
efficiencies and background rates and rejection, it is important to 
include nuclear effects in the theoretical predictions, and 
worthwhile to study kinematic distributions such as $dR/dE_r$.

\begin{figure}[th]
\begin{center}
\includegraphics[height=8cm]{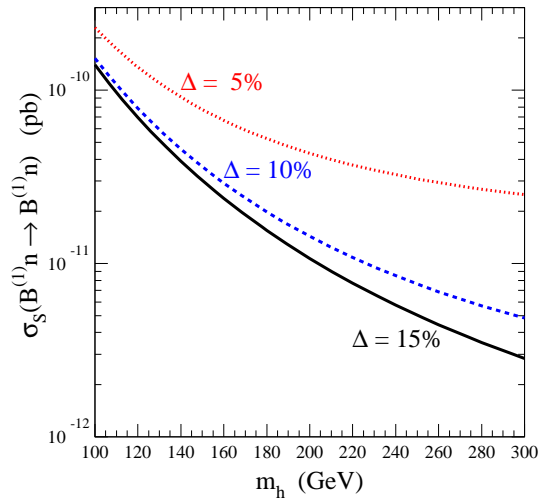}
\end{center}
\caption[]{Scalar WIMP-nucleon cross section as a function of the Higgs mass 
for  $m_{\bone}=1$ TeV and (top to bottom)
$\Delta=(m_{q^{(1)}}-m_{\bone})/m_{\bone}=5,10,15 \%$.}
\label{fig:nucleonmh}
\end{figure}

\begin{figure}[th]
\begin{center}
\includegraphics[height=8cm]{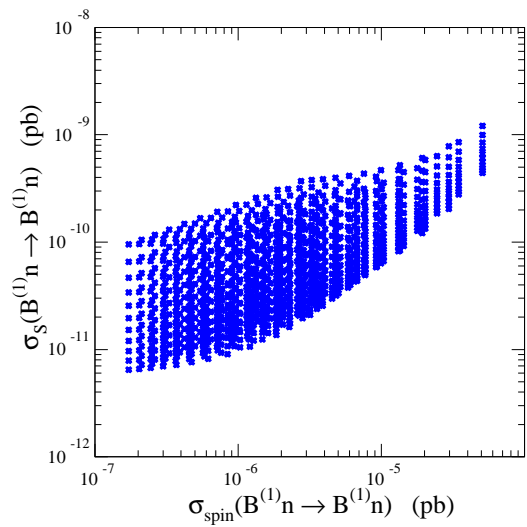}
\end{center}
\caption[]{Predictions for $\bone$-nucleon cross sections in the 
spin-independent -- spin-dependent plane. We have varied three 
parameters: $m_{\bone}$ in the 600-1200 GeV range, $\Delta$ in
the 5-15 $\%$ range and $m_h$ in the 100-200 GeV range.}
\label{fig:SIversusSD}
\end{figure}

\clearpage
\section{Differential and Integrated Event Rates}

From Eq.~(\ref{eq:dRdE}), the number of events per kilogram of detector 
per keV per day is proportional to, 
\be
\frac{dR}{dE_r} \propto \frac{\sigma_0}{m \mu^2}F^2(|\bfq|).
\ee
Larger $m$ and smaller $\sigma_0$ combine with a suppression 
from $F^2(|\bfq|)$, making the event rate quite low. The rates are
further suppressed by nuclear form factors which drop quickly as the 
recoil energy increases.

\begin{table}[th]
\begin{center}
\begin{tabular}{lcc}
\mbox{Nucleus}-$A$  & \mbox{Typical Recoil Momentum} & \mbox{Typical Recoil Energy} \\
                    & $\bfq=\mu v$                   & $|\bfq|^2/2M$ \\
& & \\
\mbox{Silicon}-23   & 22 \mbox{MeV}                  & 11 \mbox{keV} \\
\mbox{Sodium}-29    & 28 \mbox{MeV}                  & 14 \mbox{keV} \\
\mbox{Germanium}-73 & 68 \mbox{MeV}                  & 32 \mbox{keV} \\
\mbox{Iodine}-127   & 112 \mbox{MeV}                 & 50 \mbox{keV} \\
\mbox{Xenon}-131    & 116 \mbox{MeV}                 & 51 \mbox{keV}
\end{tabular}
\end{center}
\caption{Typical nucleus recoil momenta and energies after scattering with a
WIMP with mass $m \gg M$.}
\label{tab:typical}
\end{table}

In Table~\ref{tab:typical} we list some typical recoil momenta and
energies (corresponding to a WIMP mass of 1 TeV and
velocity of $v \sim 220$ km/s $\sim 10^{-3}c $) scattering from various nuclei.
Note that there is effectively a maximal recoil energy which is 
roughly 16 times the typical recoil energy listed in the table, because
the maximum velocity is approximately the galactic escape velocity,
$v_{esc}=650\pm200$ km/s $\sim 2 v$ and $q_{max}=2\mu v_{max}$.  However, at such energies the 
nuclear form factor itself already provides a high suppression
in the differential rate, such that one arrives at a good approximation 
to the integrated rate by integrating up to an energy which is four 
times the typical energy.  Thus, it is enough for our purposes to 
present $dR/dE_r$ over a 200 keV range of recoil energy. 
Experimentally, it may be useful to look at energies $\gsim 100 $ keV for 
which we expect the background to fall off. 
We illustrate the importance of nuclear effects in 
Fig.~\ref{fig:FormFactorSuppression} where we plot the scalar and spin 
form factors for $\i127$, as a function of the WIMP mass. We can see that 
they lead, for a 1 TeV WIMP, to a suppression of the cross section by a 
factor of approximately 15.

\begin{figure}[th]
\begin{center}
\includegraphics[height=8cm]{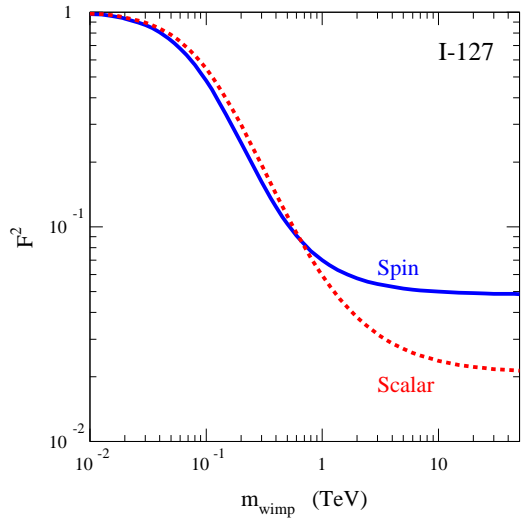}
\end{center}
\caption[]{Spin and scalar nuclear form factors for Iodine as a function of the WIMP mass. 
Note that form factors are usually expressed as a function of the recoil energy $E_r={\bf
q}^2/2M$. To generate this plot, we chose a typical recoil momentum ${\bf q}=\mu v$
so that 
$E_r(m)=m^2v^2M/(2(m+M)^2)$.}
\label{fig:FormFactorSuppression}
\end{figure}

We now examine the differential rate with respect to recoil energy for
several materials.  This distribution is important in order to correctly
apply experimental efficiencies as well as to assess signal-to-background
levels as a function of the cut on the recoil energy, $E_r^{min}$.
In Fig.~\ref{fig:Diff_Rate} we present the predictions for rates differential
in recoil energy on three different targets: NaI, $\GE73$, and $\xe131$,
including both spin-dependent and scalar contributions along with
appropriate nuclear effects.
For scalar contributions, this is the form factor given in Eq.~(\ref{eq:SFF}).
For spin-dependent form factors, we use those presented in 
\cite{Dimitrov:1994gc} for $\GE73$.
For Iodine, Sodium and $\xe131$, we adopt the spin form factors of 
\cite{Ressell:1997kx} estimated by considering the Nijmegen (II) 
nucleon-nucleon potential.

\begin{figure}[th]
\begin{center}
\includegraphics[height=9cm]{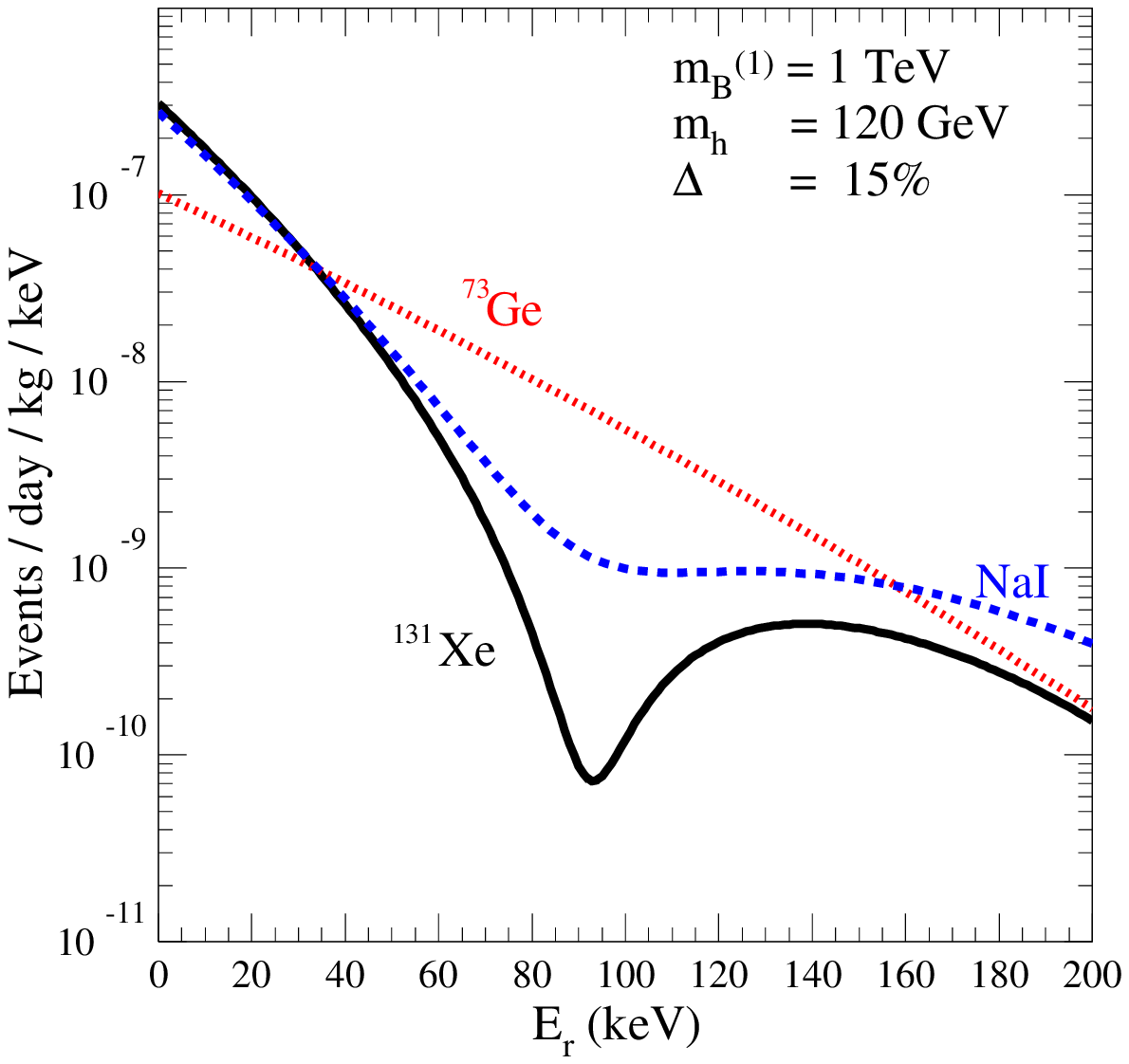}
\end{center}
\caption[]{Energy Spectrum of events for three types of detectors: 
$\GE73$ (dotted curve), NaI (dashed curve) and $\xe131$ (solid curve)
for $m_{\bone}=1$ TeV, $m_h$=120 GeV and $\Delta=15\%$.}
\label{fig:Diff_Rate}
\end{figure}

\begin{table}[th]
\begin{center}
\begin{tabular}{lccc}
Experiment & Target   & Mass         & $E_r^{min}$ \\
\\
DAMA       & NaI                & 100 kg       & 20 keV       \\
DAMA/LIBRA & NaI                & 250 kg       & 20 keV       \\
\\
GENIUS     & $\GE73$            & 100 kg       & 11 keV      \\
GENIUS II  & $\gE76$, $\gEe74$  & 100-10000 kg & 11 keV      \\
MAJORANA   & $\gE76$, $\gEe74$  & 500 kg       & 11 keV      \\
\\
XENON      & $\xe131$           & 1000 kg      & 4 keV       \\
\end{tabular}
\end{center}
\caption{Present and near-future dark matter detection experiments.}
\label{tab:expts}
\end{table}

In order to obtain the observable rates at detectors, we integrate the
differential cross sections over the recoil energies to which they are
sensitive.  The minimum observable energy is a function of the experimental
set-up and background levels.  In Table~\ref{tab:expts} we present
some current and near-future dark matter search experiments, including
their primary target nucleus, target mass, and an estimation
of the minimum recoil energy required for an observable event.
For the NaI detectors, we have included DAMA (100 kg of NaI)
and LIBRA (an upgrade of DAMA: 250 kg of NaI) \cite{Bernabei:2002ae}.  There are also
several different detectors based on various isotopes of Germanium.
The first stage of GENIUS is composed of 100 kg of $\GE73$, whereas
the second stage consists of between 100-10000 kg of a mixture
of $86\%$ $\gE76$ and $14\%$ $\gEe74$ \cite{Klapdor-Kleingrothaus:2000eq}.
The MAJORANA experiment will search for double-beta-decay with 
500 kg of the same mixture of $86\%$ $\gE76$ and $14\%$ $\gEe74$ 
\cite{Aalseth:2002sy}.
Finally, the proposed XENON experiment will consist of 1000 kg of
$\xe131$ \cite{Aprile:2002ef}.

\begin{figure}[th]
\begin{center}
\includegraphics[height=12cm]{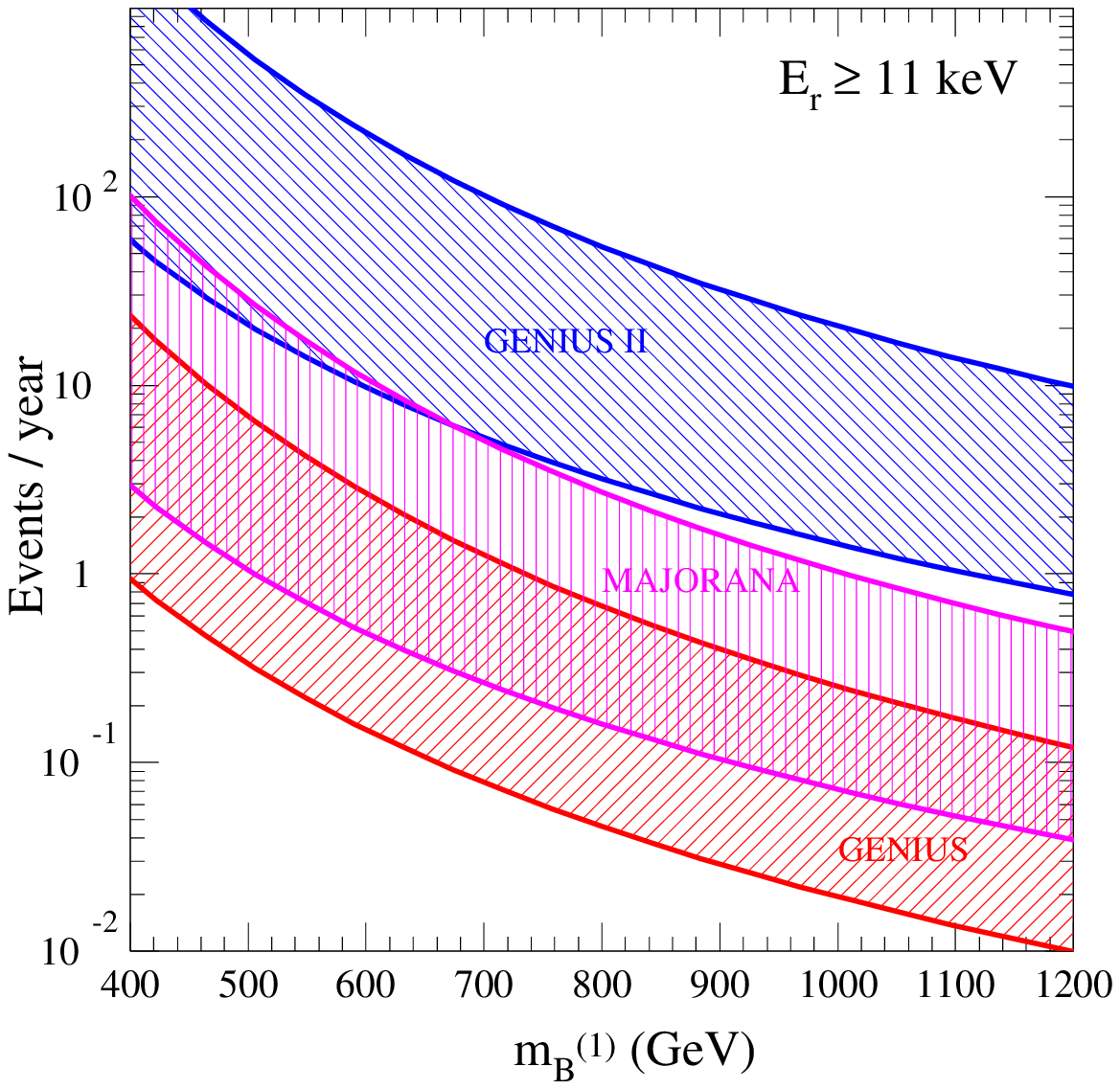}
\end{center}
\caption[]{Number of events per year for the 100 kg $\GE73$ Genius experiment,
the $10^4$ kg $\gE76$/$\gEe74$ GENIUS stage II, and 500 kg 
$\gE76$/$\gEe74$ MAJORANA experiment.  The bands are obtained by varying
$5\% \leq \Delta \leq 15\%$ and $115$ GeV $\leq m_h \leq 200$ GeV.}
\label{Genius100kg}
\end{figure}

In Fig.~\ref{Genius100kg} we show the potential number of events per year
at detectors based on Germanium isotopes, assuming $E^{min}_r$ is
11 keV.  The bands represent potential signal rates as a function of the
LKP mass, varying $5\% \leq \Delta \leq 15 \%$ and 
$115$ GeV $\leq m_h \leq 200$ GeV.  To illustrate the importance of large
mass detectors, we have chosen to compare the 100 kg $\GE73$ GENIUS, 
$10^4$ kg $\gE76$ and $\gEe74$ GENIUS stage II, and 500 kg
$\gE76$ and $\gEe74$ MAJORANA experiments.  Evident from the figure,
in order to have even one event per year at GENIUS or MAJORANA, $m_h$
and $\Delta$ must be on the small side of the band, and/or $m_{\bone}$
must be less than about 1 TeV.  In order to have ten or more events per year,
at MAJORANA, we must have $m_{\bone} \leq 700$ GeV as well as small
$m_h$ and $\Delta$.  Thanks to its enormous mass, the upgraded 
GENIUS experiment with $10^4$ kg of $\gE76$ and $\gEe74$ can do much 
better, with more than ten events per year when $m_{\bone} \leq 700$ GeV even for
unfavorable $m_h$ and $\Delta$, and more than ten events over
the entire range of $m_{\bone}$ for optimal $m_h$ and $\Delta$.

\begin{figure}[th]
\begin{center}
\includegraphics[height=12cm]{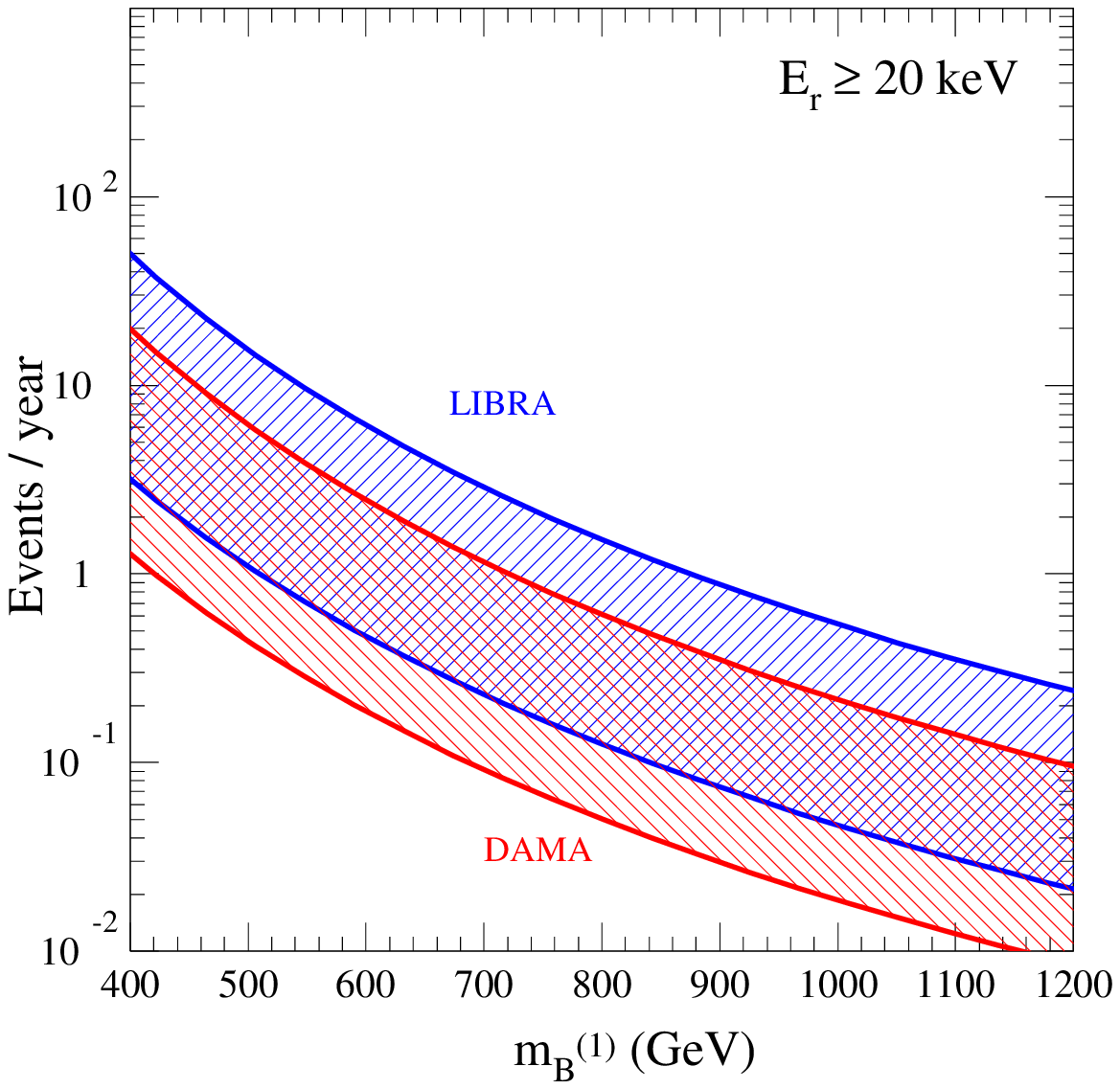}
\end{center}
\caption[]{Number of events per year for the 100 kg NaI DAMA experiment,
and 250 kg NaI DAMA/LIBRA experiment.  The bands are obtained by varying
$5\% \leq \Delta \leq 15\%$ and $115$ GeV $\leq m_h \leq 200$ GeV. Note that a signal in a
DAMA-like experiment is not given by the number of WIMP-nucleus scattering events; rather, it
is the annual modulation of this event rate, which amounts to a small fraction of the signal
rate. Experiments of this kind use a different methodology as they do not
attempt to distinguish between signal and background on an event-by-event
basis. Our figure should then be interpreted as a theorist prediction of the number of events 
100 kg and 250 kg NaI detectors can detect per year. In the particular case of the DAMA/LIBRA
experiment, the amplitude of the modulation must scale like the total rate.}
\label{NaI_year}
\end{figure}

In Fig.~\ref{NaI_year} we show events per year, varying the parameters as 
above at DAMA and DAMA/LIBRA.  DAMA, with 100 kg of NaI can observe more than ten 
events per year if the LKP is light and parameters favorable.  
 DAMA/LIBRA, with 250 kg of NaI can observe the lightest relevant LKP masses
even when $\Delta$ and $m_h$ are at the larger end we consider.
Finally, the XENON experiment combines a heavy $\xe131$ target
with a large 1000 kg detector.  The estimated events per year are plotted
in Fig.~\ref{xenon}, and are comparable to the end-stage of GENIUS II with
$10^4$ kg of Germanium.  For comparable masses, XENON could in fact do better,
thanks to the heavier target nucleus.

Our results indicate that to directly detect Kaluza-Klein dark matter,
heavy target nuclei and large mass detectors are essential.  Of course, the
actual reach of the experiments will depend on experimental issues such
as efficiencies and backgrounds, which are beyond the scope of this work.
However, given the relatively large event rates which are possible at planned
experiments, further detailed study of this subject is warranted. 

\begin{figure}[th]
\begin{center}
\includegraphics[height=12cm]{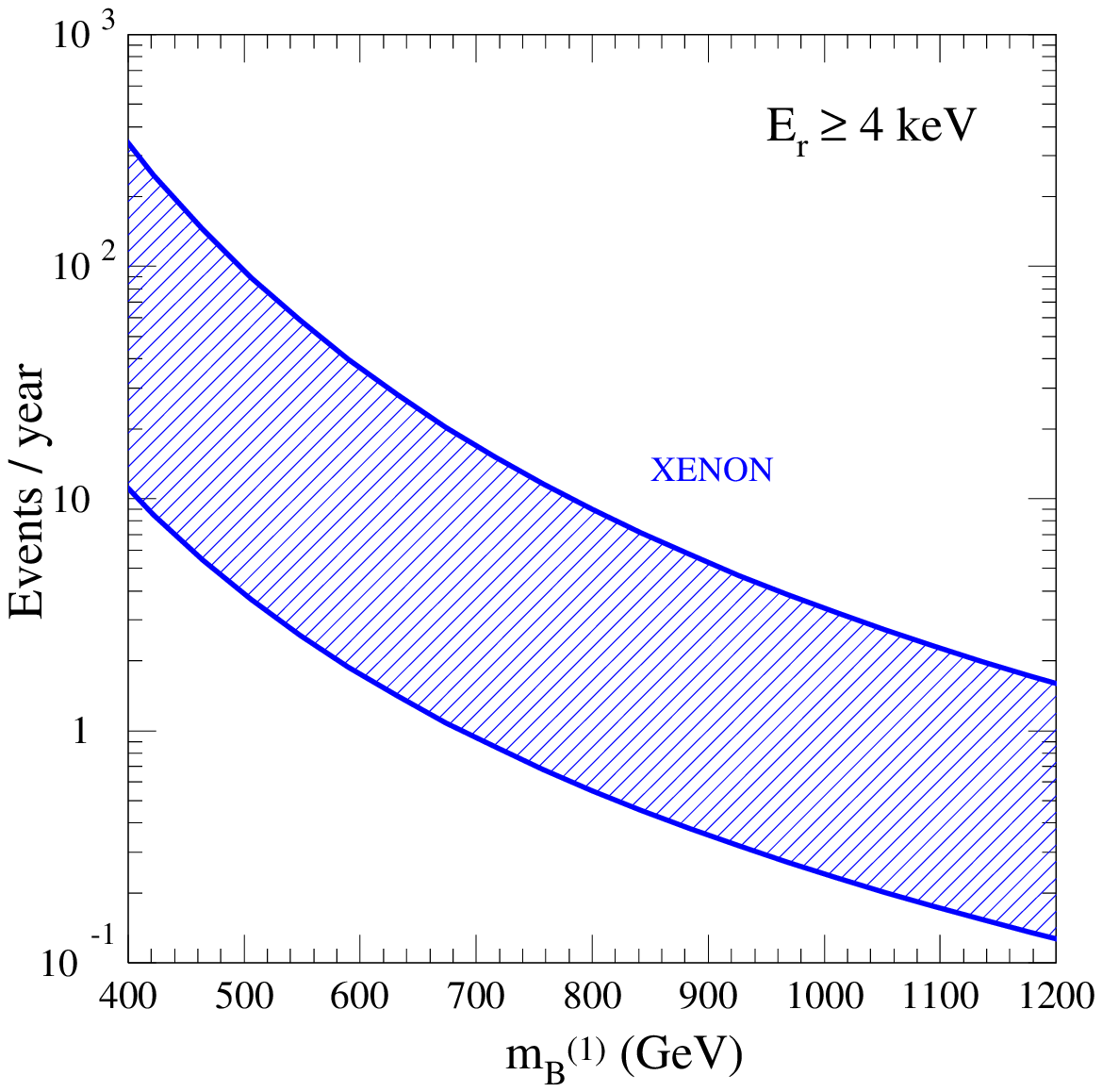}
\end{center}
\caption[]{Number of events per year for the 1000 kg $\xe131$ XENON 
experiment.  The bands are obtained by varying
$5\% \leq \Delta \leq 15\%$ and $115$ GeV $\leq m_h \leq 200$ GeV.}
\label{xenon}
\end{figure}

\section{Conclusion}
\label{sec:conclusion}

The identity of the dark matter is one of the most intriguing puzzles
in modern physics, and has sparked a major experimental program to search
for these elusive objects.  In fact, one of the primary motivations of the
supersymmetric standard model (with $R$-parity) is the fact that it 
has a natural WIMP candidate.  However, we have recently seen that a large
class of extra-dimensional models with a KK parity also provides a natural
WIMP - the heavy Kaluza-Klein modes of the ordinary photon and $Z$ boson.
While KK parity is not a necessary feature of models with extra dimensions
(just as $R$-parity is an unnecessary feature of models with supersymmetry),
it can be imposed self-consistently in the low energy effective theory.
Estimates of the relic density indicate that such particles should have masses
at the TeV scale, at the frontier of current collider and direct detection
searches.

In this article we have made a detailed study of the direct detection 
of LKPs which scatter off of heavy nuclei.  Subject to our assumptions that
boundary terms are small (perhaps generated radiatively) and
common for all quarks, our predictions depend on three parameters:
the LKP mass, the splitting between the LKP and the first level KK quarks,
and the mass of the zero-mode Higgs.  The mixture of KK $B$ and $W_3$ 
in the LKP is another parameter, which would also be interesting to study
in more detail.  This situation is to be contrasted with the minimal
supersymmetric standard model, which requires four parameters to describe the
neutralino alone.
We find that nuclear effects are important to include 
in the cross sections, particularly because of the heavy WIMP masses
favored by the estimates of the relic density.  These effects render
rates at current dark matter searches small, and indicate that future
experiments composed of large numbers of heavy nuclei can study most, but not
all of the parameter space relevant for the correct WIMP relic density.

Another search strategy which we have not considered here
relies on indirect detection, in which WIMPs annihilate one another 
producing energetic photons and positrons \cite{Cheng:2002ej,Bertone:2002ms} or
neutrinos \cite{Cheng:2002ej,Hooper:2002gs,Bertone:2002ms} which can be observed on the Earth.
The resulting LKP masses which can be probed by next-generation experiments
are similar to those determined here from planned direct searches such
as GENIUS.

In conclusion, Kaluza-Klein dark matter is well-motivated in a large class
of theories with compact extra dimensions, and provides an interesting 
alternative 
to the standard neutralino LSP in a supersymmetric model.  Accurate 
determination of scattering cross sections with heavy nuclei involve nuclear
form factors, and indicate that future experiments can study a significant
region of parameter space.  At the same time, collider searches at the 
Tevatron Run II and LHC will study a similar range of parameter space.
The exciting scenario of KK dark matter can be studied on two fronts 
simultaneously in the near future.

\section*{Acknowledgments}

We are grateful to John Beacom, Hsin-Chia Cheng, Juan Collar and Neil Weiner 
for discussions. We thank  the hospitality of the Aspen Center for Physics 
as well as the T-8 group of Los Alamos National laboratory 
for the Santa Fe Institute 2002 where part of 
this work was completed.  This work is supported in part by the US 
Department of Energy, High Energy Physics Division, under contract 
W-31-109-Eng-38 and also by the David and Lucile Packard Foundation.
Fermilab is operated by Universities Research
Association Inc. under contract no. DE-AC02-76CH02000 with the DOE.


\end{document}